\documentclass{PoS}

\usepackage{graphicx}
\usepackage{dcolumn}
\usepackage{bm}
\usepackage{epstopdf}
\usepackage{mathrsfs}
\usepackage{amssymb}
\usepackage{amsmath}

\def\del{\partial}

\def\pp{\bar{p}}
\def\pb{\bar {\boldsymbol p}}

\def\bkappa{{\boldsymbol \kappa}}
\def\bbkappa{\bar{\boldsymbol \kappa}}

\def\sM{\text{med}}

\def\pk{p\cdot k}
\def\hk{\bar{p} \cdot k}
\def\kh{{\boldsymbol \kappa}}
\def\khb{\bar{{\boldsymbol \kappa}}}
\def\bl{\boldsymbol{L}}
\def\bbl{\bar{\boldsymbol{L}}}

\newcommand{\beq}{\begin{equation}}
\newcommand{\eeq}{\end{equation}}
\newcommand{\bes}{\begin{subequations}}
\newcommand{\ees}{\end{subequations}}
\newcommand{\bal}{\begin{align}}
\newcommand{\eal}{\end{align}}
\newcommand{\be}{\begin{eqnarray}}
\newcommand{\ee}{\end{eqnarray}}

\title{In-medium antenna radiation in $t$-channel}

\ShortTitle{In-medium antenna radiation in $t$-channel}

\author{N\'{e}stor Armesto, \speaker{Hao Ma}, Mauricio Mart\'{i}nez, and Carlos A. Salgado\\
        Departamento de F\'isica de Part\'iculas and IGFAE,
Universidade de Santiago de Compostela \\
E-15782 Santiago de Compostela, 
Galicia-Spain}
        
\author{Yacine Mehtar-Tani\\
        Institut de Physique Th\'eorique, CEA Saclay, 
F-91191 Gif-sur-Yvette, France}

\abstract{The color coherence effects in the medium modification to the initial state radiation are studied via a simple setup which allows these effects to be pronounced. The medium-induced gluon radiation spectrum off a hard quark which suffers a highly virtual photon scattering and subsequently crosses a dilute QCD medium of finite size is obtained. The angular distribution of the medium-induced gluon radiation spectrum is modified when the interference contributions between the incoming and outgoing quarks at a finite scattering angle are included.}

\FullConference{Sixth International Conference on Quarks and Nuclear Physics,\\
		April 16-20, 2012\\
		Ecole Polytechnique, Palaiseau, Paris}

\begin{document}

\section{Introduction}
\label{sec:intro}
Most of the efforts made in recent years have concentrated on the medium modification for the inclusive gluon spectrum off a single hard parton \cite{Baier:1996kr,Baier:1996sk,Zakharov:1996fv,Zakharov:1997uu,Wiedemann:2000za,gyu00}. One clear limitation of these formalisms is the lack of coherence effects between different emitters which is a fundamental ingredient for the construction of the parton shower in vacuum. Recently, an interesting program is introduced to address the problem of coherence effects in the presence of a QCD medium from first principles \cite{MehtarTani:2010ma,MehtarTani:2011tz,MehtarTani:2011jw,CasalderreySolana:2011rz}.  These groups studied  the case of gluon emission off two emitters, a $q\bar{q}$ antenna formed initially in either a color singlet or a color octet state for the massless \cite{MehtarTani:2010ma,MehtarTani:2011gf} and the massive case \cite{Armesto:2011ir} in a dilute QCD medium as well as for a dense medium in the soft limit \cite{MehtarTani:2011tz} and finite gluon energies \cite{MehtarTani:2011jw,CasalderreySolana:2011rz}. The results of these works have shown an intriguing and interesting structure due to interference effects in a QCD medium. In this contribution, we study the medium modifications to the initial state radiation in a dilute QCD medium.

\section{Vaccuum emission pattern of the initial state radiation}
\label{sec:vacinit}
In vacuum, gluon emissions from the space-like cascades is usually called {\it initial state radiation}. In this section we briefly describe the vacuum emission pattern of the initial state radiation. In the classical limit, the inclusive gluon spectrum with $4$-momentum $k=(\omega,\vec{k})$ is given by
\beq
\label{eq:incluspec}
(2\pi)^3 2\omega \ \frac{dN}{d^3\vec{k}}=\sum_{\lambda=\pm 1}|{\cal M}^a_\lambda(\vec{k})|^2 \,.
\eeq
The scattering amplitude is obtained from the classical gauge field by the reduction formula:
\beq
\label{eq:redform}
{\cal M}^a_\lambda (\vec{k})=\lim_{k^2\to 0 } -k^2 A^a_\mu (k) \epsilon^\mu_\lambda \,.
\eeq
The classical gauge field $A_\mu\equiv A^a_\mu t^a$ (being $ t^a$ the generator of $SU(3)$ in the fundamental representation) is the solution of the classical Yang-Mills (CYM) equations $[D_\mu,F^{\mu\nu}]=J^\mu$. The covariant derivative is  defined as $D_\mu=\partial_\mu - ig A_\mu$ and the non-Abelian field strength tensor is $F_{\mu\nu}=\partial_\mu A_\nu-\partial_\nu A_\mu -ig [A_\mu,A_\nu]$. The projectiles are described by a classical eikonalized current $J^\mu$ which is covariantly conserved, i.e. $[D_\mu,J^\mu]=0$. Our calculation is performed in the light cone gauge $A^+=0$. By doing this, it must be understood that only the physical, transverse polarizations of the gluon contribute to the cross section, and hence $\sum_\lambda \epsilon^{i \, *}_\lambda \epsilon^j_\lambda = \delta^{ij}$ ($i,j=1,2$). The classical eikonalized current that describes the incoming and outgoing quarks with the photon scattering at certain time $t=t_0$ reads $J^\mu_{(0)}= J_{in}^{\mu,a}+J_{out}^{\mu,a}$, where the currents for the incoming and outgoing quarks read
\bes
\label{eq:inoutcurs}
\be
J_{in,a}^{\mu}&=& g\frac{p^\mu}{p^+}\,\delta^{(3)}\biggl(\vec{x} - \frac{\vec{p}}{E} t\biggr) \Theta (t_0-t) Q^{in}_{a}\,,\\
J_{out,a}^{\mu}&=& g\frac{\pp^\mu}{\pp^+}\,\delta^{(3)}\biggl(\vec{\bar x} - \frac{\vec{\pb}}{\bar E} t\biggr) \Theta (t-t_0) Q^{out}_{a}\,,
\ee
\ees
where $Q^{in (out)}_{a}$ is the color charge of the incoming (outgoing) quark. By current conservation, one has $Q^{in}_{a}=Q^{out}_{a}$, and therefore $Q^2_{in,a}=Q^2_{out,a}=Q_{in,a}\cdot Q_{out,a}=C_F=\bigl(N_C^2 -1\bigr)/2N_C$.  Due to the choice of the gauge, it is suitable to perform the calculation in the light cone variables, i.e. $k\equiv \bigl[k^+=(\omega+k^3)/\sqrt{2},k^-=(\omega-k^3)/\sqrt{2},{\bf k} \bigr]$, ${\bf k}=(k^1,k^2)$. By linearizing the CYM equations, the solution of the classical gauge field at leading order in momentum space reads
\beq
\label{eq:solvac}
-k^2 A_{(0)}^{i,a}=\, 2ig \left( \frac{\kappa^i}{\bkappa^2}Q_{in}^a - \frac{\bar\kappa^i}{\bbkappa^2}Q_{out}^a\right) \,.
\eeq
In the last expression we have introduced the transverse vector $\kappa^i = k^i - x\,p^i$ (similar definition goes for the outgoing quark), which describes the transverse momentum of the gluon relative to the momentum of the incoming (outgoing) quark. In addition, we define the momentum fractions carried out by the emitted gluon as $x= k^+/p^+$ and $\bar{x}=\bar{k}^+/\pp$, respectively.

Taking the solution of the classical gauge field into the reduction formula and summing over the physical polarizations, one can show that in vacuum the inclusive gluon spectrum is
\beq
\label{vacspec}
\omega\frac{dN^\text{vac}}{d^3\vec{k}}=\frac{\alpha_s  C_F}{(2\pi)^2}\,
\frac{2\,p\cdot\pp}{(\pk)\,(\hk)}
\equiv \frac{\alpha_s\,C_F}{(2\pi)^2\,\omega^2}\,\big( \mathcal{R}_{in} + \mathcal{R}_{out} - 2 \mathcal{J} \big)  \,,
\eeq
where we have defined the independent contribution of the incoming quark as $\mathcal{R}_{in}=4\omega^2/\bkappa^2$ and analogously for the outgoing quark. The interference between the two emitters corresponds to the term $\mathcal{J}=4\omega^2\,\kh\cdot\khb/\bigl(\bkappa^2\,\bbkappa^2\bigr)$. Eq. (\ref{vacspec}) presents two types of divergences, when the emitted gluon is soft ($\omega\to 0$) and when it is emitted collinearly to either the incoming or the outgoing quark. Notice that the inclusive gluon spectrum is suppressed at large angles due to destructive interference effects between the emitters. This result can be understood intuitively from first principles: if the transverse wavelength of the emitted gluon is larger than the transverse displacement of the color current of the quark, the radiated gluon is unable to resolve it, and hence soft gluon radiation at large angles is suppressed. The suppression of large angle emission is better understood by performing an average over the azimuthal angles of either the incoming or outgoing quark. In order to separate the collinear divergences of either the incoming or outgoing quark, we define $\mathcal{P}_{in}=\mathcal{R}_{in}-\mathcal{J}$ and $\mathcal{P}_{out}=\mathcal{R}_{out}-\mathcal{J}$, and hence $\mathcal{R}_{sing}=\mathcal{P}_{in}+\mathcal{P}_{out}$. This separation of the radiation pattern allows us to interpret each of these terms as the coherent gluon emission off the quark. For instance, if we choose the momentum of the incoming quark along the longitudinal $z$-axis and perform the azimuthal angle average, we obtain
\beq
\label{eq:azivacin}
\int_0^{2\pi}\,\frac{d\varphi}{2\pi}\,\mathcal{P}_{in}= \frac{2}{1-\cos\theta}\,\Theta (\cos\theta-\cos\theta_{qq})\,.
\eeq
Similar result is found if one makes the azimuthal angle average over the outgoing quark. Eq. (\ref{eq:azivacin}) shows that gluon emission is confined in the cone with opening angle $\theta_{qq}$ along the incoming or outgoing quark. This property is known as {\it angular ordering}.  The corresponding gluon emission probability off the incoming quark in vacuum reads
\beq
\label{eq:incvacspec}
\langle dN^{vac}_{in} \rangle_{\varphi} = \frac{\alpha_s \,C_F}{\pi}\,\frac{d\omega}{\omega}\, \frac{\sin\theta d\theta}{1-\cos\theta}\Theta\bigl(\cos\theta - \cos\theta_{qq}\bigr)\,,
\eeq
likewise for the gluon radiation off the outgoing quark. The procedure described above can be extended to higher orders and is the basic building block for the construction of a coherent parton branching formalism.

\section{Medium modifications to the initial state radiation}
\label{sec:medgluon}
To include medium modifications to the radiation pattern of the initial state, we consider a highly energetic quark produced in the remote past which suffers a photon scattering at $x^+_0$ and afterwards it crosses a static dilute QCD medium of finite size $L^+= L/\sqrt{2}$, being $L$ the size of the medium. The quark starts its interaction with the QCD medium exactly at $x^+=x^+_0$. We allow to have gluon emissions either before reaching the QCD medium or once the quark is crossing the QCD medium (see the figure in \cite{t-channel}). In the semiclassical approach, the quark field acts as a perturbation around the strong medium field $A_{med}$ and the total field is written as
\beq
\label{eq:totA}
A^\mu \equiv A^\mu_{med}  + A^\mu_{(0)} + A^\mu_{(1)}\,,
\eeq
where $A_{(0)}$ is the gauge field of the quark in vacuum and $A_{(1)}$ is the response of the field at first order in $A^\mu_{med}$. The medium gauge field is described by $A^-_{med}(x^+,\bold{x})$ and it is a solution of a two-dimensional Poisson equation $-\boldsymbol{\partial} A^-_{med}(x^+,\bold{x})=\rho (x^+,\bold{x})$, where $\rho (x^+,\bold{x})$ is the static distribution of medium color charges, which is treated as a Gaussian white noise. Notice that in this approximation, one has $A^i_{med}=A^+_{med}=0$. In Fourier space the medium gauge field reads
\beq
\label{eq:Amed}
A^-_{med} (q)=\,2\pi\,\delta(q^+)\,\int_{x^+_0}^\infty dx^+\,\mathcal{A}^-_{med}(x^+,\bold{q})\,e^{i\,q^-x^+}\,.
\eeq
At first order in the medium field, the continuity relation for the induced eikonalized current reads $\partial_\mu J^\mu_{(1)}=\,ig\,\bigl[A^-_{med},J^+_{(0)}\bigr]$. Its solution can be written as
\be
\label{eq:totindcurr}
J^\mu_{(1)}&=&ig\frac{p^\mu}{p\cdot \del }~ \bigl[A^-_{\sM}, J^+_{in}\bigr] + ig \frac{ \bar{p}^\mu}{ \bar{p}\cdot \del }~ \bigl[A^-_{\sM}, J^+_{out}\bigr] \,\nonumber\\
&\equiv& J^\mu_{in,(1)} +J^\mu_{out,(1)}\,,
\ee
where $J^+_{in (out)}$ are given by Eqs. (\ref{eq:inoutcurs}). The current $J^\mu_{in,(1)}$ for the incoming quark is
\beq
\label{eq:incurrsol}
J^\mu_{in,(1)}(k)=(ig)^2\frac{p^\mu}{-i(p\cdot k) }\int\!\! \frac{d^4q}{(2\pi)^4}~\frac{p^+}{p\cdot (k-q)-i\epsilon}~i[T\cdot A^-_{\sM}(q)]^{ab} Q_{in}^b \,,
\eeq
where $[T\cdot A^-_{\sM}(q)]^{ab} Q_{in}^b \equiv -i f^{abc}A^c_{\sM}(q)Q^b_{in}$ and $f^{abc}$ is the $SU(3)$ structure constants. A similar expression is found for the current $J^\mu_{out,(1)}$ for the outgoing quark. After linearizing the CYM equations, the equation of motion for the transverse components of the induced gauge field $A^i_{(1)}$ is
\beq
\label{eq:A1eq}
\square A_{(1)}^i-2ig  \bigl[A^-_{\sM}, \, \del^+ A_{(0)}^i\bigr] = -\frac{\del^i}{\del^+}J_{(1)}^++J_{(1)}^iÊ\,.
\eeq
The solution of Eq. (\ref{eq:A1eq}) in Fourier space for the induced gauge field of the incoming quark is written as
\beq
\label{eq:Ain1sola}
-k^2 A^-_{in,(1)} = 2g \int \frac{d^4q}{(2\pi)^4}\, (k-q)^+ \bigl[A_{\sM}^-(q), A^i_{in,(0)}(k-q)\bigr] -\frac{k^i}{k^+}J^+_{in,(1),a}+J^i_{in,(1),a} \,,
\eeq
where $A^i_{in,(0)}(k)$ is identified with the incoming quark-induced vacuum field from Eq. (\ref{eq:solvac}):
\beq
-k^2A^{i,a}_{in,(0)}(k)\;=\;- 2ig\frac{\kappa^i}{\bkappa}\,Q_{in}^a \;.
\eeq
Integrating out $q^-$ in Eq. (\ref{eq:Ain1sola}) and imposing the condition that the medium starts at $x^+=x^+_0$, one obtains
\beq
\label{eq:Ain1solb}
-k^2A_{in,(1)}^{i,a} = 2ig^2  \int \frac{d^2 {\bf q}}{(2\pi)^2}\,\int^{+ \infty}_{x^+_0} d x^+\, [T\cdot A^-_{\sM}(x^+,{\bf q})]^{ab} Q_{in}^b\,e^{i \left(k^--\frac{({\bf k}-{\bf q})^2}{2 k^+}\right)x^+} \frac{(\kappa-q)^i}{\bigl(\bkappa-{\bf q}\bigr)^2}  \,.
\eeq
By following a similar procedure one gets the solution for the induced gauge field of the outgoing quark:
\be
\label{eq:Aout1sol}
-k^2A_{out,(1)}^{i,a}& = &- 2ig^2  \int \frac{d^2 {\bf q}}{(2\pi)^2}\,\int^{+ \infty}_{x^+_0} d x^+\, [T\cdot A^-_{\sM}(x^+,{\bf q})]^{ab} Q_{out}^b\, e^{i \left(k^--\frac{({\bf k}-{\bf q})^2}{2 k^+}\right)x^+} \nonumber \\
&& \left[\frac{(\bar{\kappa}-q)^i}{(\bbkappa-{\bf q})^2}\left\{1-\exp\biggl(i \frac{({\bf k}-{\bf q})^2}{2 k^+}x^+\biggr)\right\}+\frac{\bar{\kappa}^i}{\bbkappa^2} \exp\biggl(i \frac{({\bf k}-{\bf q})^2}{2 k^+}x^+\biggr) \right] \,.
\ee
Thus, by using the reduction formula together with Eqs. (\ref{eq:Ain1sola}) and (\ref{eq:Aout1sol}) one gets the total scattering amplitude for gluon radiation off the incoming and outgoing quarks:
\be
\label{eq:totmedampl}
{\cal M}^a_{\lambda} &=& {\cal M}^a_{\lambda,in} +{\cal M}^a_{\lambda,out}\,\nonumber\\
&=&2ig^2  \int \frac{d^2 {\bf q}}{(2\pi)^2}\,\int^{L^+}_{0} d x^+\, [T\cdot A^-_{\sM}(x^+,{\bf q})]^{ab}\nonumber\\
&&\times\biggl\{
Q_{in}^b\,\frac{\bkappa-{\bf q}}{\bigl(\bkappa-{\bf q}\bigr)^2} 
-Q_{out}^b\, \left[\frac{\bar{\bkappa}-{\bf q}}{(\bbkappa-{\bf q})^2}-\bbl \exp\biggl(i \frac{({\bf k}-{\bf q})^2}{2 k^+}x^+\biggr) \right] \biggr\}\,,
\ee
where we set $x_0^+=0$. In addition, we use the definition of the transverse component of the Lipatov vertex in the light cone gauge:
\beq
\label{eq:lipvertex}
\bbl=\frac{\bar{\bkappa}-{\bf q}}{(\bbkappa-{\bf q})^2}-\frac{\bar{\bkappa}}{\bbkappa^2}.
\eeq
The contribution to the total scattering amplitude from the incoming and outgoing quarks is not the same. Such difference arises due not only to the kinematic constraints but also to the fact that the incoming quark is created in the remote past where there is no medium.

To evaluate the cross section we must average over the medium field and include the virtual corrections, the so-called contact terms \cite{Baier:1996kr,Baier:1996sk, Wiedemann:2000za}. To perform the medium average, we assume that the medium color charges have just transverse correlation but not  longitudinal one. Therefore the medium charge density is a Gaussian white noise and the average over the medium field can be written as
\beq
\label{eq:MediumAverage}
\langle A^-_{\sM,a}(x^+,{\bf q}) A^{- *}_{\sM,b}(x'^+,{\bf q}')\rangle = \delta^{ab} m_D^2~ n_0~\delta(x^+-x'^+)(2\pi)^2 \delta^{(2)}({\bf q}-{\bf q}') \, {\cal V}^ 2({\bf q}) \,,
\eeq
where $ {\cal V}({\bf q})=1/({\bf q}^2+m_D^2)$ is the Yukawa-type potential, $m_D$ is the Debye mass and $n_0$ is the one-dimensional density that describes the distribution of the scattering centers in the QCD medium. The contact terms required by unitarity are the interferences between the gluon emission amplitude in vacuum and the one accompanied by two-gluon exchange with the medium. These can be added to the radiative cross section through a redefinition of the potential ${\cal V}^2({\bf q})\to{\cal V}^2({\bf q})-\delta^2({\bf q})\int d^2{\bf q}' {\cal V}^2({\bf q}')$. Squaring the amplitude and summing over the polarization vectors, $|{\cal M}|^2= |{\cal M}|^2_{in}\,+\,|{\cal M}|^2_{out}\,+\,2\text{Re}{\cal M}_{in}{\cal M}^*_{out}$, the spectrum of the medium-induced gluon radiation is
\be
\label{eq:totmedspec}
\omega\frac{dN^\text{med}}{d^3\vec{k}} &=&\frac{4\,\alpha_s C_F \,\hat q}{\pi} \int \frac{d^2 {\bf q}}{(2\pi)^2}{\cal V}^2({\bf q})~\int_0^{+ \infty} dx^+\,\Biggl[\frac{1}{(\bkappa-{\bf q})^2}-\frac{1}{\bkappa^2}\nonumber\\
&&+\,2\,\frac{\bbkappa\cdot{\bf q}}{\bbkappa^2(\bbkappa-{\bf q})^2}\biggl(1-\cos\biggr[\frac{({\bf k}-{\bf q})^2}{2 k^+}x^+\biggr]\biggr)\,\nonumber\\
&&-\,2\,\Biggl\{\bl \cdot \frac{\bbkappa}{\bbkappa^2}\,+\,
\bbl\cdot\frac{(\bkappa-{\bf q})}{(\bkappa-{\bf q})^2}\,
\biggl(1-\cos\biggr[\frac{({\bf k}-{\bf q})^2}{2 k^+}x^+\biggr]\biggr)\,\Biggr\}
\Biggr],
\ee
where we define $\hat{q}=\alpha_s C_A n_0 m_D^2$. The novel contributions associated to the interferences between both emitters are contained in the third line of Eq. (\ref{eq:totmedspec}). The interferences depend on two terms, the interferences between the emission currents associated to the Lipatov vertex $\bl$ and the independent radiation component of the outgoing quark, and the interferences between the emission current $\bbl$ and the radiation component of the incoming quark which rescatters inside the medium. Notice that the interference terms have different phase structures. This is a consequence of our setup since the contributions of the rescattering of the incoming quark get cancelled inside the medium. Both interference terms are soft and collinear divergent. The medium-induced gluon radiation spectrum has an interesting asymptotic limit. For the case when the opening angle between both emitters $\theta_{qq}\to 0$, Eq. (\ref{eq:totmedspec}) reduces to the well known  Gunion-Bertsch (GB) spectrum \cite{Gunion:1981qs}. This is expected since GB spectrum is the induced gluon emission due to the scattering of an asymptotic quark with the medium.

\end{document}